\documentclass[runningheads,a4paper]{llncs}

\usepackage{amssymb}
\usepackage{amsmath}
\usepackage{graphicx}
\usepackage{subfigure}
\usepackage{algorithm}
\usepackage{algorithmic}
\usepackage{tabularx, booktabs}
\newcolumntype{Y}{>{\centering\arraybackslash}X}
\newcolumntype{L}[1]{>{\raggedright\let\newline\\\arraybackslash\hspace{0pt}}m{#1}}

\usepackage{url}
\urldef{\mailsa}\path|{ehsan.amid, onur.dikmen, erkki.oja}@aalto.fi|

\newcommand{\y}{\ensuremath{\mathbf{y}}}
\newcommand{\p}{\ensuremath{\mathbf{p}}}
\newcommand{\q}{\ensuremath{\mathbf{q}}}
\newcommand{\Dkl}{\ensuremath{D_{\text{KL}}}}
\newcommand{\betap}{{\beta+1}}

\newcommand{\betapp}{({\beta+1})}

\begin{document}

\mainmatter  

\title{Optimizing the Information Retrieval Trade-off in Data Visualization Using $\alpha$-Divergence}

\titlerunning{Optimizing the Information Retrieval Trade-off}

%
%
\author{Ehsan Amid \and Onur Dikmen \and Erkki Oja}
\authorrunning{Amid et al.}

\institute{Department of Computer Science, Aalto University,\\
02150 Espoo, Finland\\
\mailsa\\
}


%
%

\toctitle{Optimizing the Information Retrieval Trade-off}
\tocauthor{Optimizing the Information Retrieval Trade-off}
\maketitle

\begin{abstract}

Data visualization is one of the major applications of nonlinear dimensionality reduction. From the information retrieval perspective, the quality of a visualization can be evaluated by considering the extent that the neighborhood relation of each data point is maintained while the number of unrelated points that are retrieved is minimized. This property can be quantified as a trade-off between the mean precision and mean recall of the visualization. While there have been some approaches to formulate the visualization objective directly as a weighted sum of the precision and recall, there is no systematic way to determine the optimal trade-off between these two nor a clear interpretation of the optimal value. In this paper, we investigate the properties of $\alpha$-divergence for information visualization,  focusing our attention on a particular range of $\alpha$ values. We show that the minimization of the new cost function corresponds to maximizing a geometric mean between precision and recall, parameterized by $\alpha$. Contrary to some earlier methods, no hand-tuning is needed, but we can rigorously estimate the optimal value of $\alpha$ for a given input data. For this, we provide a statistical framework using a novel distribution  called Exponential Divergence with Augmentation (EDA). By the extensive set of experiments, we show that the optimal value of $\alpha$, obtained by EDA corresponds to the optimal trade-off between the precision and recall for a given data distribution.

\end{abstract}

\section{Introduction}
\label{sec:intro}

Dimensionality reduction and, in particular, data visualization has been
a prominent research track for the past few decades as an important
step in data analysis. Many non-linear dimensionality reduction
methods such as Sammon mapping~\cite{Sammon}, Isomap~\cite{isomap},
Locally Linear Embedding~\cite{LLE}, Maximum Variance Unfolding \cite{MVU}, and Laplacian Eigenmaps~\cite{Laplacian} have been proposed to overcome the
shortcomings of the simple linear methods in handling data lying on
non-linear manifolds. Although all these methods have been
successfully applied to several artificial as well as real-world
datasets, they have been particularly designed for unraveling the underlying low-dimensional manifold rather than providing a good visualization for the user. More specifically, these methods fail to represent the data properly on a two-dimensional display, especially when the inherent dimensionality of the manifold is higher than two.

Preserving the neighborhood structure of the data points in a two-dimensional visualization is crucial for gaining initial intuition about the data. From an information retrieval perspective, the fundamental trade-off in a good visualization is to minimize the number of missing related points in the neighorhood of each point while avoiding the unrelated points to appear in the neighborhood. In other words, the goal of a good visualization is to provide a faithful representation of the data by achieving the best possible balance between \emph{precision} and \emph{recall}. Venna \textit{et al}.~\cite{nerv} impose the trade-off between precision and recall by penalizing the violation of each by a different cost and then, obtain the final visualization by minimizing the total cost. Their method, Neighborhood Retrieval Visualizer (\textsf{NeRV}), achieves different visualizations by adjusting the trade-off parameter, based on the application. Their method also covers Stochastic Neighbor Embedding (\textsf{SNE}) method~\cite{sne} as special case of maximizing the recall in the visualization. However, the trade-off parameter does not have any statistical interpretation and hence, must be set manually, by the user. Additionally, the \textit{optimal trade-off} between the precision and recall for a particular dataset is unknown.

In this paper, we consider the problem of finding the visualization of the data which achieves the optimal trade-off between the precision and recall. For this, we use a more general class of divergences, called $\alpha$-divergence, as the cost function. This choice leads to a wider range of values, including the cost function of \textsf{NeRV}. For estimating the optimal value of $\alpha$,
we present a statistical framework based on a recently proposed
distribution called Exponential Divergence with Augmentation
(\textsf{EDA})~\cite{onur}. \textsf{EDA} is an approximate generalization of Tweedie
distribution, which has a well-established relation to
$\beta$-divergence. With a nonlinear transformation, an equivalence
between $\beta$ and $\alpha$-divergences can be shown and \textsf{EDA} can also
be used for estimation of $\alpha$. As our contributions, we provide a proof that minimizing the $\alpha$-divergence is equivalent to maximizing the geometric mean between precision and recall, parameterized by $\alpha$. This provides an upper bound for the trade-off, achieved by \textsf{NeRV}. By an extensive set of experiments on different types of datasets, we also show that the visualization obtained by using the optimal value of $\alpha$ provides a faithful representation of the original data, attaining the optimal trade-off between precision and recall. 

The organization of the paper is as follows. We first start with
briefly introducing the information retrieval perspective for data visualization in Section~\ref{sec:problem} and
then, provide the motivation for using the $\alpha$-divergence as the cost function and explore the characteristics of its gradient. In Section~\ref{sec:opt}, we present our framework for estimating the optimal value of $\alpha$ for a given
data distribution. We provide our experimental results in
Section~\ref{sec:experiments} and finally, draw conclusions in
Section~\ref{sec:conclusion}.

\section{Problem Definition}
\label{sec:problem}

\subsection{Information Retrieval Perspective for Dimensionality Reduction}
\label{subsec:info_ret}

Let $\{\mathbf{x}_i\} \in \mathbb{R}^D$ denote the high-dimensional representation of the input points. 
The \emph{binary neighbors} of point $i$ in the input space, denoted by $P_i$, can be defined as the set of points that fall within a fixed radius or a fixed number of nearest-neighbors of $\mathbf{x}_i$. 
From an information retrieval perspective, the aim of the dimensionality reduction is to provide the user a low-dimensional representation of the data, i.e. $\{\y_i\} \in \mathbb{R}^d$, in which the neighborhood relation of the points is preserved as much as possible. In other words, the number of relevant points that are retrieved for each point is maximized while minimizing the number of  irrelevant points appearing appearing in the neighborhood. These notions can be quantified by considering the mean \emph{precision} and mean \emph{recall} of the embedding. For this purpose, let $Q_i$ denote the binary neighbors of each point in the low-dimensional embedding, defined in a similar manner. The precision and recall for point $i$ is defined as the number of points that are common in $P_i$ and $Q_i$ divided by the size of $P_i$ and $Q_i$, respectively. The mean precision and mean recall of the embedding can be calculated by talking the average over the precision and recall values of all points, respectively.


Venna et al.~\cite{nerv} introduce a generalization of the binary neighborhood relationship by providing a probabilistic model of neighborhood for each point. In other words, each point $j\neq i$ has a non-negative probability $q_{ij}$ of being a relevant neighbor of point $i$ in the embedding. This probability can be obtained by normalizing a non-increasing function of distance between $i$ and $j$ in the embedding. Similar probabilistic neighborhood relation $p_{ij}$ can be defined in the input space. Under the probabilistic models of neighborhood, a natural way to obtain the embedding is to minimize the sum of Kullback-Leibler (KL) costs between $p_{ij}$'s and $q_{ij}$ values. This is essentially the cost function adopted in the \textsf{SNE} method~\cite{sne}, that is,
\begin{equation}
\label{eq:C_sne}
\mathcal{C}_{\text{SNE}}= \sum_i \Dkl(\mathbf{p}_i \Vert \mathbf{q}_i) = \sum_i \sum_j p_{ij} \log \frac{p_{ij}}{q_{ij}}\, .
\end{equation}
It is shown in~\cite{nerv} that  $\Dkl(\mathbf{p}_i\Vert\mathbf{q}_i)$ is a generalization of recall. Therefore, the \textsf{SNE} method is equivalent  to maximizing the recall in the visualization. On the other hand, $\Dkl(\mathbf{q}_i\Vert\mathbf{p}_i)$ induces a generalization of precision. Thus, the \textsf{NeRV} method ~\cite{nerv} promotes using a convex sum of $\Dkl(\mathbf{p}_i\Vert\mathbf{q}_i)$ and $\Dkl(\mathbf{q}_i\Vert\mathbf{p}_i)$ divergences as the cost function, 
\begin{equation}
\label{eq:C_nerv}
\mathcal{C}_{\text{NeRV}}= \lambda \sum_i \Dkl(\mathbf{p}_i \Vert \mathbf{q}_i) +  (1- \lambda) \sum_i  \Dkl(\mathbf{q}_i|\mathbf{p}_i)\, .
\end{equation}
Parameter $0 \leq \lambda \leq 1$ controls the trade-off between maximizing the generalized precision ($\lambda = 1$) or maximizing the generalized recall ($\lambda = 0$) in the embedding. Under the binary neighborhood assumption, these become equivalent to maximizing precision and recall, respectively.

While imposing a balance between precision and recall in a visualization happens to be crucial, the \textsf{NeRV} method induces the extra cost of setting the parameter $\lambda$ to a reasonable value. The user might try several different values of $\lambda$ and asses the results manually. However, there is no systematic way to estimate the optimal value using the data. To overcome this problem, we introduce a more general class of divergences, called $\alpha$-divergence, which includes the \textsf{NeRV} cost function as special cases, in the following section. 

\subsection{$\alpha$-Divergence for Stochastic Neighbor Embedding}
The (asymmetric) $\alpha$-divergence over discrete distributions is defined by
\begin{equation}
D_{\alpha}(\mathbf{p}\Vert \mathbf{q}) = \frac{\sum_i p_i^{\alpha} q_i^{1-\alpha}-\alpha p_i + (\alpha-1)q_i}{\alpha(\alpha-1)}\,.
\end{equation}
As special cases, it includes many well-known divergences, such as
$\Dkl(\mathbf{q} \Vert \mathbf{p})$ and $\Dkl(\mathbf{p}\Vert \mathbf{q})$ which are
obtained in the limit $\alpha\rightarrow 0$ and $\alpha\rightarrow 1$,
respectively~\cite{NMF}. We are mainly interested in the interval $\alpha \in [0,1]$. The points $\alpha
= 0$ and $\alpha = 1$ amount to the cost function of \textsf{NeRV} when
$\lambda = 0$ and $\lambda = 1$, respectively. More generally, when
$\alpha$ varies from $0$ to $1$, $\alpha$-divergence passes smoothly
through all values of \textsf{NeRV} cost function for $\lambda \in (0,1)$ since
the divergence itself is a continuous function of $\alpha$. Note that the mapping from $\lambda$ to $\alpha$ is not onto; this can be easily
seen by considering two arbitrary distributions and varying $\lambda$
and $\alpha$ and, finding the value of each. Thus, $\alpha$-divergence
covers an even wider range of values compared to the convex sum of $\Dkl(\mathbf{p}\Vert \mathbf{q})$ and $\Dkl(\mathbf{q} \Vert \mathbf{p})$.

As an important result, $\alpha$-divergence amounts to maximizing a geometric mean of precision and recall, parameterized by $\alpha$. As an immediate result of the geometric-arithmetic mean inequality, it can be shown that maximizing the geometric mean exceeds the maximum arithmetic mean for equal trade-off parameters $\alpha = \lambda$. The proof is in Appendix~A.

The aforementioned properties promote investigating a sum of $\alpha$-divergences over all pairs of distributions as the new cost function. We call the method \textsf{$\alpha$-SNE}, for Stochastic Neighbor Embedding with $\alpha$-divergence. The new cost function to minimize becomes
\begin{equation}
\label{eq:C_asne}
C_{\text{$\alpha$-SNE}} = \sum_i D_{\alpha}(\mathbf{p}_i\Vert \mathbf{q}_i) = \left\{ 
  \begin{array}{l l}
    \sum_i \sum_j \displaystyle{\frac{p_{ij}^{\alpha} q_{ij}^{1-\alpha}-\alpha p_{ij} + (\alpha-1)q_{ij}}{\alpha(\alpha-1)}} & \quad \alpha \neq 0,1\\\\
    \sum_i \sum_j q_{ij} \log(q_{ij}/p_{ij}) & \quad \alpha = 0\\\\
   \sum_i \sum_j p_{ij} \log(p_{ij}/q_{ij}) & \quad \alpha = 1
  \end{array} \right. \,.
\end{equation}

\subsection{Notes on the Gradient and Practical Matters on Optimization}

More interesting properties are revealed by considering the gradient, 
\begin{equation}
\label{aSNE_grad}
\frac{\partial C_{\alpha\text{-SNE}}}{\partial y_i} = 
\frac{2}{\alpha}\sum_{j\neq i} (y_i - y_j) \left(p_{ij}^\alpha q_{ij}^{1-\alpha}-\theta_i q_{ij} + p_{ji}^\alpha q_{ji}^{1-\alpha}-\theta_j q_{ji}\right), \quad \alpha \neq 0\,,
\end{equation}
in which $0 \leq \theta_i = \sum_{j\neq i} p_{ij}^\alpha
q_{ij}^{1-\alpha}$ and we call it the \emph{compatibility factor} for point $i$, with the following properties: $\theta_i = 1$ if $\p_i = \q_i$, and $\theta_i \leq 1$ for $\alpha \in (0,1]$ (except $\alpha = 0$\footnote{The gradient for the case $\alpha = 0$ can be obtained in the limit $\alpha \rightarrow 0$ where we have
\begin{equation*}\frac{\partial C_{\alpha\text{-SNE}}}{\partial y_i} = 2\sum_{j\neq i} (y_i - y_j)
\left(q_{ij}D_{\text{KL}}(\mathbf{q}_i \Vert \mathbf{p}_i) -
  q_{ij}\log\frac{q_{ij}}{p_{ij}} +  q_{ji}D_{\text{KL}}(\mathbf{q}_j\Vert \mathbf{p}_j)
  - q_{ji}\log\frac{q_{ji}}{p_{ji}}\right).
  \end{equation*}}). The gradient has an
interpretation of springs between map points with stiffness proportional
to the mismatch in the probability distributions, similar to SNE,
\begin{equation}
\label{SNE_grad}
\frac{\partial C_{\text{SNE}}}{\partial y_i} = 2\sum_{j}(y_i-y_j)\left(p_{ij}-q_{ij}+p_{ji}-q_{ji}\right)\,.
\end{equation}
However, comparing
the gradient with the gradient of \textsf{SNE}, 
it can be seen
that the attraction terms $p_{ij}$ and $p_{ji}$ are replaced by
$p_{ij}^\alpha q_{ij}^{1-\alpha}$ and $p_{ji}^\alpha
q_{ji}^{1-\alpha}$, respectively. On the other hand, the repulsion
terms $q_{ij}$ and $q_{ji}$ are weighted by the compatibility factors
for points $i$ and $j$, respectively. Therefore, the compatibility factor
for point $i$ can also be seen as the sum of the attraction terms between
$i$ and the rest of the points. Finally, the whole gradient is scaled by a
factor of $1/\alpha$. If $\alpha = 1$, then (\ref{SNE_grad}) and
(\ref{aSNE_grad}) agree.  

\begin{figure*}[t!]
        \centering
        \subfigure[Gradient of SNE]{
                \includegraphics[width=0.32\textwidth]{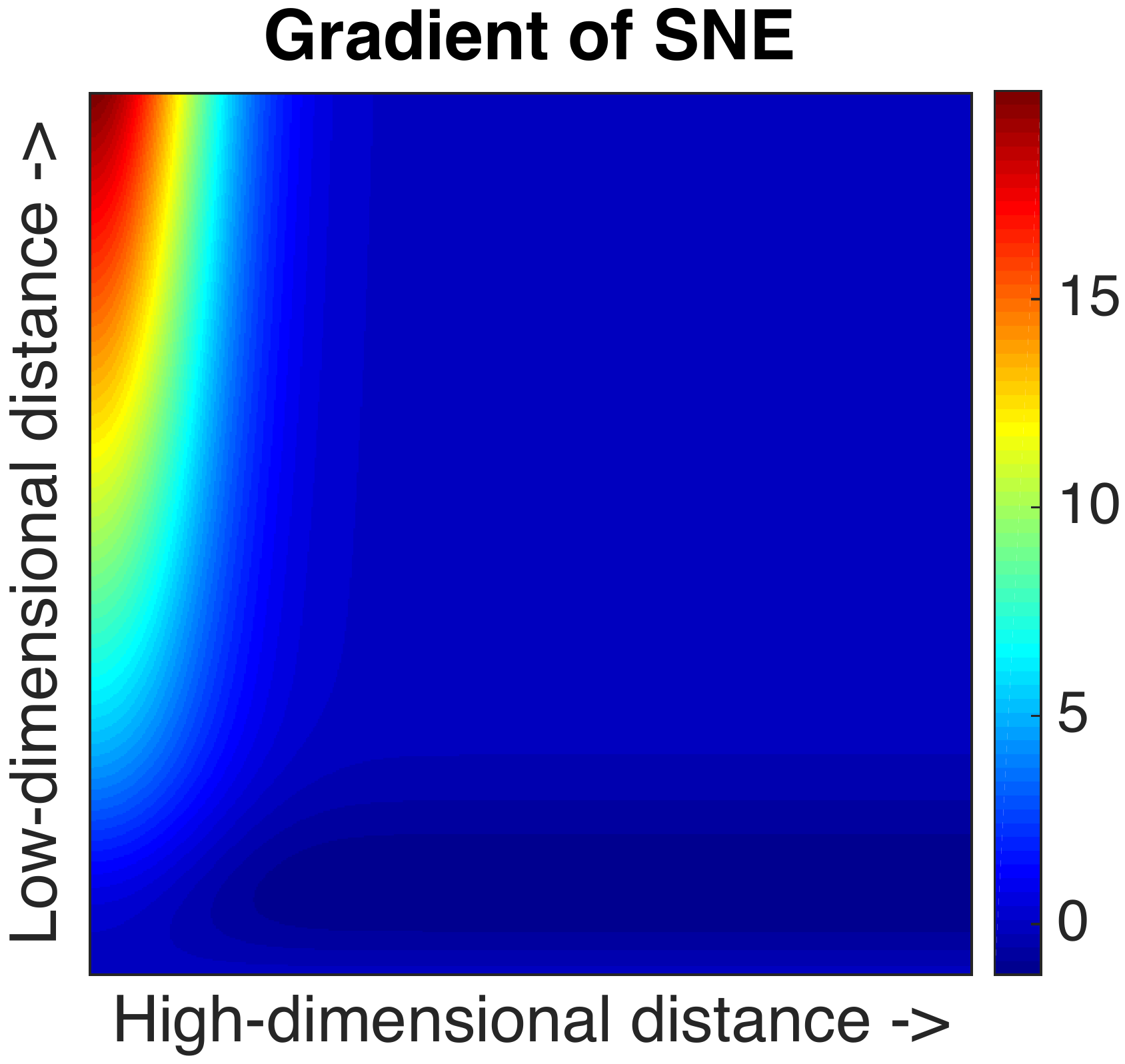}
                \label{fig:SNE_grad}}
        \subfigure[Gradient of $\alpha$-SNE]{
                \includegraphics[width=0.33\textwidth]{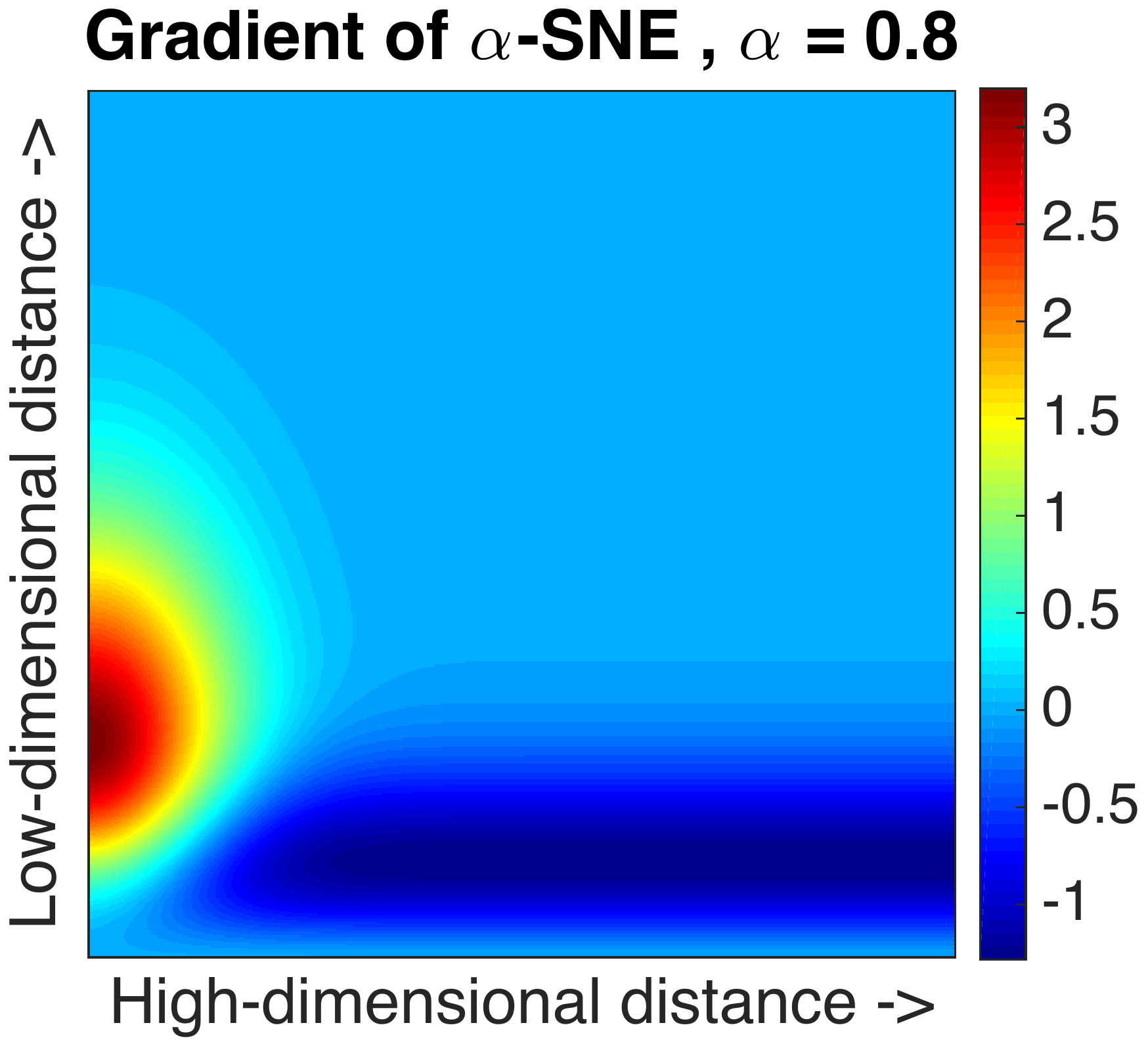}
                \label{fig:alphaSNE_grad}}
        \caption{Gradients of \textsf{SNE} and \textsf{$\alpha$-SNE} as a function of
  the pairwise Euclidean distances between two points in the
  high-dimensional space and the low-dimensional image. \textsf{$\alpha$-SNE} produce more balanced gradients
  compared to \textsf{SNE}. Note different color scales.}\label{fig:gradient}
\end{figure*}


Figure~\ref{fig:SNE_grad} and~\ref{fig:alphaSNE_grad} show the gradients of \textsf{SNE} and \textsf{$\alpha$-SNE} (with $\alpha=0.8$), respectively, for a pair of points,
as a function of their Euclidean distances in the high-dimensional
space and the low-dimensional space. 
Positive values represent
attraction, while negative values correspond to repulsion. As can be
seen in Figure~\ref{fig:SNE_grad},  \textsf{SNE} exerts a large attraction force for
moderately close datapoints which are mapped far from each other. However,
the repulsion force is comparatively small for the opposite case
(around 19 to 1). These properties are consistent with the fact that the \textsf{SNE} method maximizes the recall by collapsing the neighboring relevant points together. 
On the other hand, \textsf{$\alpha$-SNE} (Figure~\ref{fig:alphaSNE_grad}) results in a more balanced
gradient, compared to \textsf{SNE}, by damping the large attraction forces and
further, strongly repelling dissimilar datapoints which are mapped
close together. Thus, \textsf{$\alpha$-SNE} yields a more balanced trade-off between precision and recall, compared to\textsf{ SNE}.

As a final remark, the optimization of the cost function~(\ref{eq:C_asne}), for a given $\alpha$,  can be achieved using standard
methods, e.g., steepest descent. A jitter noise with a
constant variance can be used to model simulated annealing in early
stages.

\section{$\alpha$-Optimization}
\label{sec:opt}

After defining the cost function of \textsf{$\alpha$-SNE} and obtaining a
method to appropriately optimize the cost function, there remains the
problem of selecting the $\alpha$ value for a particular
dataset, which results in the optimal trade-off between precision and recall.

The optimization of the $\alpha$ parameter is performed using
Exponential Divergence with Augmentation (\textsf{EDA})~\cite{onur}, a
distribution proposed initially for maximum likelihood estimation of $\beta$ in
$\beta$-divergence $D_\beta(u||\mu)$, where $u$ and $\mu$ are any positive
vectors such as probability distributions. Typically $u$ is known and
$\mu$ is a parametric approximation. Once the optimal $\beta$ is
found, the optimal $\alpha$ is obtained by a simple transformation.  
\textsf{EDA} is an approximation to the Tweedie distribution,
$p_\mathrm{Tw}(u;\mu,\beta,\phi)$, which is related to
$\beta$-divergence in such a way that $\mu^*$ that maximizes
the likelihood of $p_\mathrm{Tw}$ also minimizes $D_\beta$. While
$\beta$-divergence does not provide a means to optimize $\beta$
directly, our basic idea is that the likelihood of $\beta$ stemming from $p_\mathrm{Tw}$ can be
maximized for that purpose. Note that both $\alpha$ and $\beta$
divergences are separable, and so we can operate component-wise. 
The pdf of the one-variate Tweedie density of $u_i$ is given as 
\begin{equation}
\label{eq:tweedie_pdf}
p_\text{Tw}(u_i;\mu_i,\phi,\beta) = f(u_i,\phi,\beta)
\exp\left[\frac{1}{\phi}\left(\frac{u_i\mu_i^\beta}{\beta}-\frac{\mu_i^\betap}{\betap}\right)\right]\,,
\end{equation}
for $\beta \neq -1, 0$ where $\phi>0$ is the dispersion parameter and $f(u_i,\phi,\beta)$ is a
weighting function, whose analytical form is generally not available
and has to be approximated. However, there are some shortcomings associated
with Tweedie likelihood, especially, the pdf does not exist for
$\beta\in (0,1)$ and approximation of $f(u_i,\phi,\beta)$ is not well
studied for $\beta>1$. The \textsf{EDA} density is proposed to overcome these
issues, while being a close approximation to Tweedie
distribution. Using the relation with $\beta$-divergence and Laplace's
method, its pdf is found to be of the form \cite{onur}
\begin{multline}
\label{eq:EDA_pdf}
p_\text{EDA}(u_i;\mu_i,\phi,\beta) =\\
\frac{1}{Z_{\beta,\phi}}\exp\left[\frac{\beta-1}{2}\log u_i  +
  \frac{1}{\phi}\left( - 
    \frac{u_i^\betap}{\beta\betapp} + \frac{u_i\mu_i^\beta}{\beta} -
    \frac{\mu_i^\betap}{\betap} \right)\right]\, ,
\end{multline}
where $Z_{\beta,\phi}$ is a normalizing constant and $D_\beta(u_i||\mu_i)$
appears in the exponent. Evaluation of $Z_{\beta,\phi}$ requires an
integration in one dimension. Although it is not available
analytically in general\footnote{In fact, $Z_{\beta,\phi}$ is
  analytically available for $\beta = 1,0,-1,-2$, which correspond to
  Gaussian, Poisson, Gamma and Inverse 
Gaussian distributions, respectively, which are also special cases of
Tweedie distribution.}, it can be evaluated numerically using standard
statistical software. The parameters $\beta$ and $\phi$ can be
optimized either by maximizing the likelihood or
using methods for parameter estimation in non-normalized densities,
such as Score Matching (\textsf{SM})~\cite{SM}. Both of these methods have been
successfully used to find optimal $\beta$ values~\cite{onur}. 

It is possible to use \textsf{EDA} to optimize $\alpha$, too, using the relation
between $\alpha$ and $\beta$-divergences. Note that both divergences
are separable and we can formulate the relation using just scalars.  We have
\begin{align}
D_\beta(u_i||\mu_i) = D_\alpha(v_i||m_i)\,,
\end{align}
with a nonlinear transformation $u_i=v_i^\alpha/\alpha^{2\alpha}$, $\mu_i=m_i^\alpha/\alpha^{2\alpha}$
and $\beta=1/\alpha-1$ for $\alpha\neq0$. This relationship allows us to evaluate the likelihood of $m_i$ and $\alpha$ using $u_i$ and $\beta$:
\begin{align}
\label{eq:trans}
p(v_i;m_i,\alpha,\phi) 
= p_\mathrm{EDA}(u_i;\mu_i,\phi,\beta) u_i^{-\beta} |\beta+1|\,.
\end{align}
\makebox[\textwidth][s]{$\alpha$ can be optimized (alongside $\phi$) by maximizing its
likelihood given by}
$p(v_i;m_i,\alpha,\phi)$ or minimizing the \textsf{SM} objective
function evaluated from above. It is more convenient to treat $m_i$ as
constant, fixed to the value which minimizes the
$\alpha$-divergence. It is also possible to optimize it using \textsf{EDA}.

To solve our original problem, we fix the $v_i$ values in~(\ref{eq:trans}) to the vectorized form of
matrix $\mathbf{P} \in \mathbb{R}^{n\times n}_+$, which contains probabilities
$P_i$ in \textsf{$\alpha$-SNE} in each column. We set $m_i$ to the vectorized
form of matrix $\mathbf{Q} \in \mathbb{R}_+^{n\times n}$ which is formed in a
similar manner for the map points. We then compute the values $u_i$ and $\mu_i$ from the above
transformation and optimize jointly over $(\alpha,\phi)$ by minimizing
the score matching objective function of the unnormalized \textsf{EDA}
density~(\ref{eq:trans}) and select the best $\alpha$ value.

\section{Experiments}
\label{sec:experiments}

\begin{table*}[t!]
	\begin{center}
	{\scriptsize
\begin{tabularx}{\textwidth}{L{19mm} *{11}{Y}}
\toprule
{\bf Dataset} & {\bf Size} & {\sf PCA} &  {\sf LLE} &  {\sf LEM} &  {\sf Isomap} &  {\sf SNE} &  {\sf t-SNE} &  {\sf HSSNE} &  {\sf NeRV} {\tiny}& {\sf $\alpha$-SNE} {\tiny} &  {\sf $\alpha$-SNE} {\tiny (EDA)}\\
\midrule
{\bf Iris} & 150 & 0.85 & 0.70 & 0.85 & 0.65 & 0.88 & 0.86 & 0.83 & 0.89 & \textbf{0.90} & 0.88 \\ 
{\bf Wine} & 178 & 0.50 & 0.49 & 0.51 & 0.47 & 0.64 & 0.69 & 0.68 & 0.69 & \textbf{0.72} & 0.69 \\ 
{\bf Image Segs$^*$} & 210 & 0.40 & 0.27 & 0.27 & 0.58 & 0.74 & 0.83 & 0.77 & 0.80 & \textbf{0.84} & 0.80 \\ 
{\bf Glass} & 214 & 0.50 & 0.50 & 0.50 & 0.53 & 0.71 & 0.73 & 0.68 & 0.73 & \textbf{0.75} & 0.74 \\ 
{\bf Leaf} & 340 & 0.63 & 0.71 & 0.68 & 0.60 & 0.71 & 0.72 & 0.66 & 0.74 & \textbf{0.76} & 0.74 \\ 
{\bf Olivetti Faces} & 400 & 0.26 & 0.24 & 0.28 & 0.25 & 0.35 & \textbf{0.46} & 0.45 & 0.44 & \textbf{0.46} & 0.44 \\ 
{\bf UMist Faces} & 565 & 0.36 & 0.32 & 0.38 & 0.49 & 0.51 & 0.72 & 0.65 & 0.70 & \textbf{0.75} & \textbf{0.75} \\ 
{\bf Vehicle} & 846 & 0.26 & 0.25 & 0.30 & 0.29 & 0.47 & 0.61 & 0.57 & 0.58 & \textbf{0.63} & \textbf{0.63} \\ 
{\bf USPS Digits$^*$} & 1000 & 0.08 & 0.06 & 0.10 & 0.16 & 0.21 & 0.38 & 0.36 & 0.34 & \textbf{0.40} & 0.39 \\ 
{\bf COIL20} & 1440 & 0.28 & 0.30 & 0.30 & 0.21 & 0.62 & 0.79 & 0.74 & 0.77 & \textbf{0.80} & 0.79 \\ 
{\bf MIT Scene} & 2686 & 0.05 & 0.04 & 0.05 & 0.07 & 0.09 & 0.21 & 0.22 & 0.19 & \textbf{0.22} & \textbf{0.22} \\ 
{\bf Texture$^*$} & 2986 & 0.27 & 0.19 & 0.28 & 0.14 & 0.49 & 0.60 & 0.55 & 0.57 & \textbf{0.62} & 0.61 \\ 
{\bf MNIST$^*$} & 6000 & 0.27 & 0.16 & 0.26 & 0.09 & 0.11 &  0.34 &   0.32 & 0.26 & {\bf 0.35} & {\bf 0.35}\\ 
\bottomrule
\multicolumn{5}{c}{{\tiny $\ast$ Only a subset of the original datasets are used.}} & & & & & & &\\
\end{tabularx}
}
\end{center}
	\caption{Area under ROC curve (AUC) for different methods.}\label{tab:results}
\end{table*}

In this section, we provide a set of experiments on different real-world datasets to asses the performance of $\alpha$-SNE compared to other dimensionality reduction methods. We consider the following methods for comparison: Principal Component Analysis ({\sf PCA}), Locally Linear Embedding ({\sf LLE}), {\sf Isomap}, Laplacian Eigenmaps ({\sf LEM}), {\sf SNE}, t-distributed SNE ({\sf t-SNE}), Heavy-tailed Symmetric SNE ({\sf HSSNE}), and {\sf NeRV}. Our implementation of {\sf $\alpha$-SNE} is in MATLAB and the code along with the description of the datasets used is available online\footnote{\url{https://github.com/eamid/alpha-SNE}}. For the other methods, we use publicly available implementations. 

As the goodness measure, we consider the area
under receiver operating characteristic (ROC) curve (AUC). In this way,  we can combine the mean precision and mean recall into a single value which can easily be used to compare the performance of different approaches.  To calculate AUC, we fix the
neighborhood size in the input space to 20-nearest neighbors and
vary the number of neighbors in the output space from 1 to 100 to
calculate precision and recall. Finally, we compute the area under the resulting ROC curve. For each dataset, we repeat the
experiments 20 times with different random initializations and report
the averages over all the trials. For {\sf LLE}, {\sf Isomap}, and {\sf LEM}, we consider a wide range of parameters and report the maximum AUC obtained. We set the tail-heaviness parameter in {\sf HSSNE} equal to $2$. For {\sf NeRV}, we perform a linear search over different values of $\lambda$ and select the one which results in maximum AUC. For our method, we consider two approaches: as the greedy approach, we perform a linear search over different values of $\alpha$ and report the maximum AUC value obtained. Moreover, we apply the optimization framework in Section~\ref{sec:opt} and report the AUC value obtained using the optimal value of $\alpha$. By this, we can also asses the performance of \textsf{EDA} for finding the optimal $\alpha$ for information retrieval. 

\begin{figure*}[t!]
        \centering
        \subfigure{
                \includegraphics[width=0.23\textwidth]{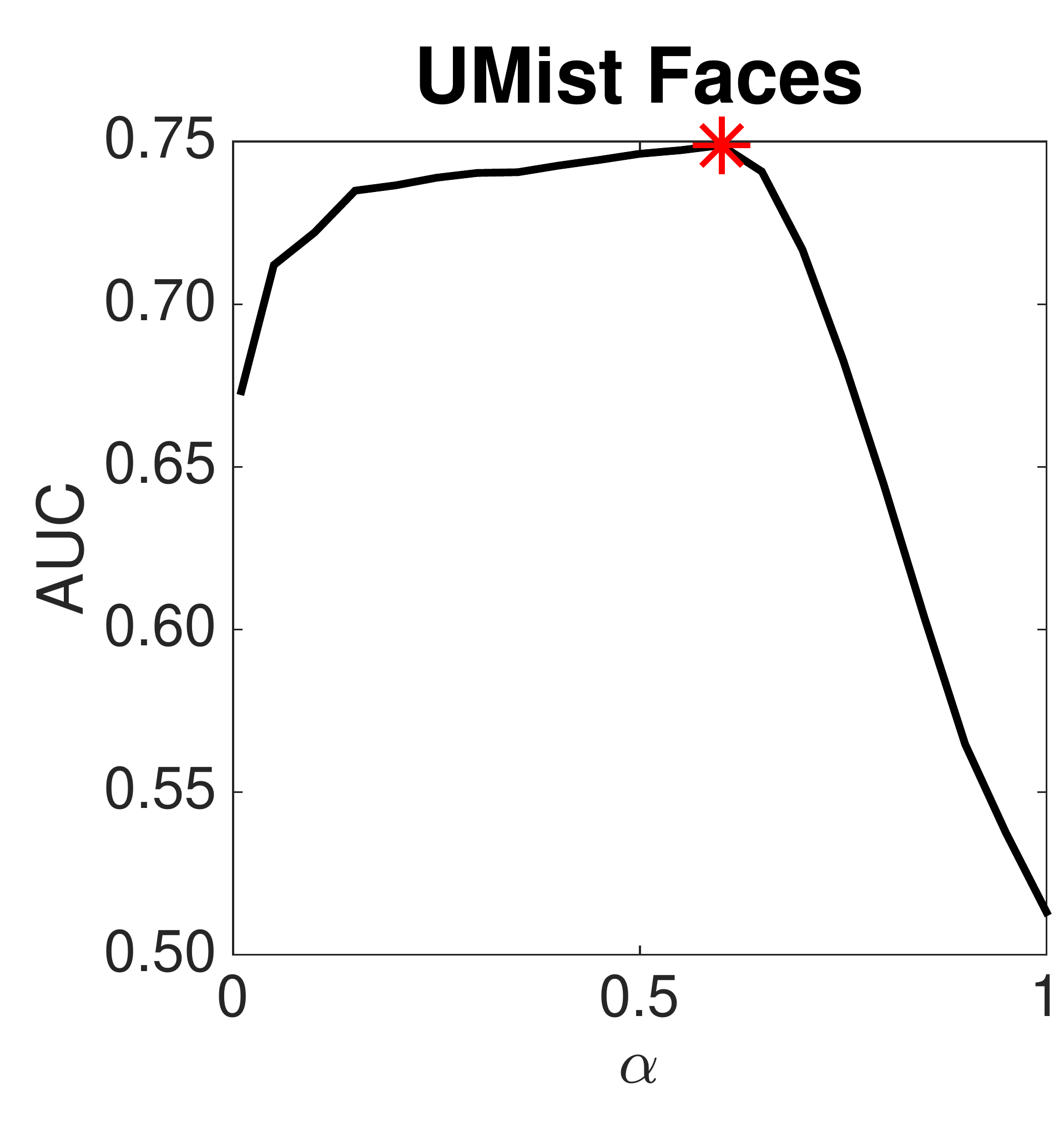}
                \label{fig:auc_ufaces}}\hfill
        \subfigure{
                 \includegraphics[width=0.23\textwidth]{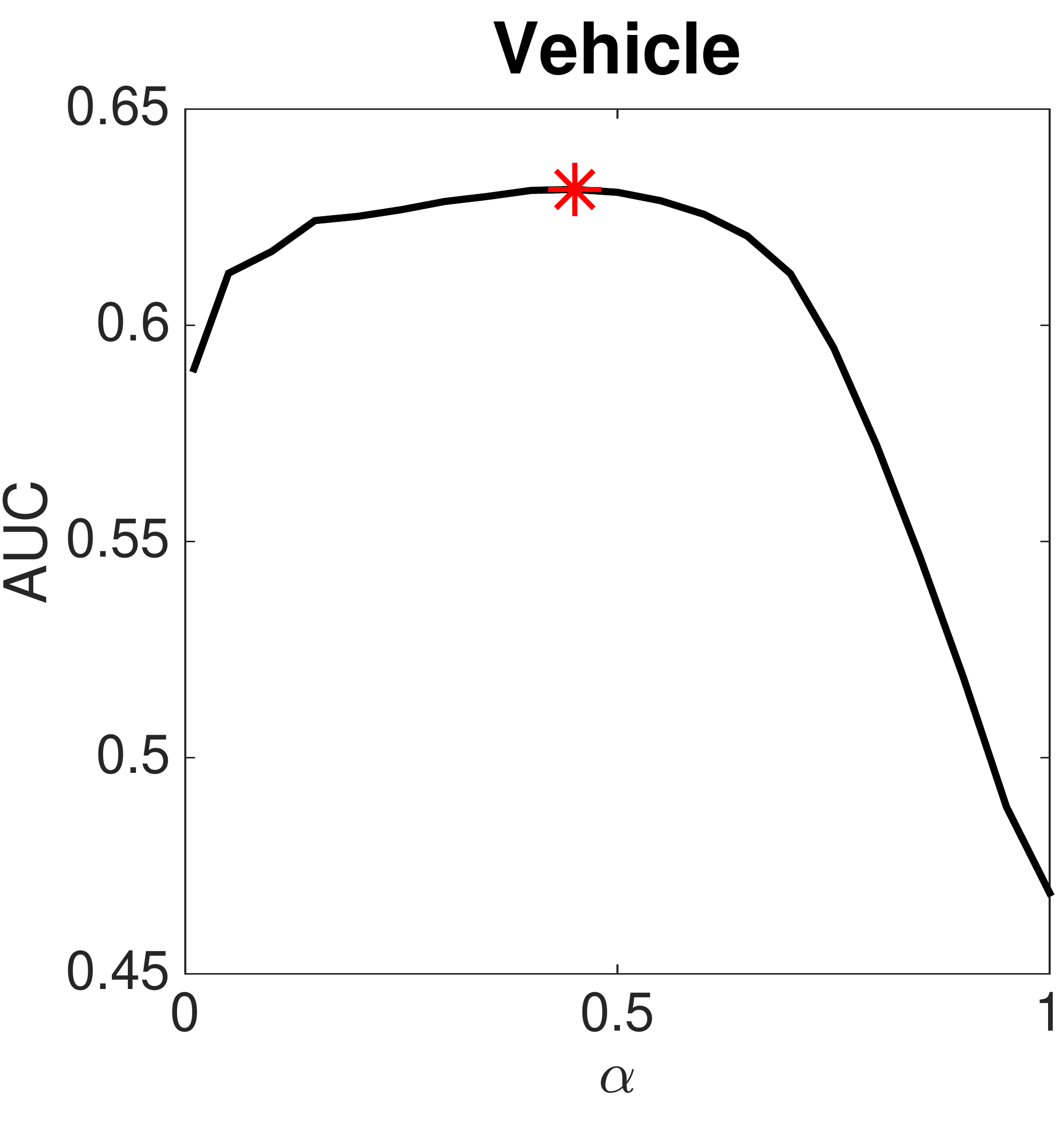}
                \label{fig:auc_vehicle}}\hfill
        \subfigure{
                 \includegraphics[width=0.23\textwidth]{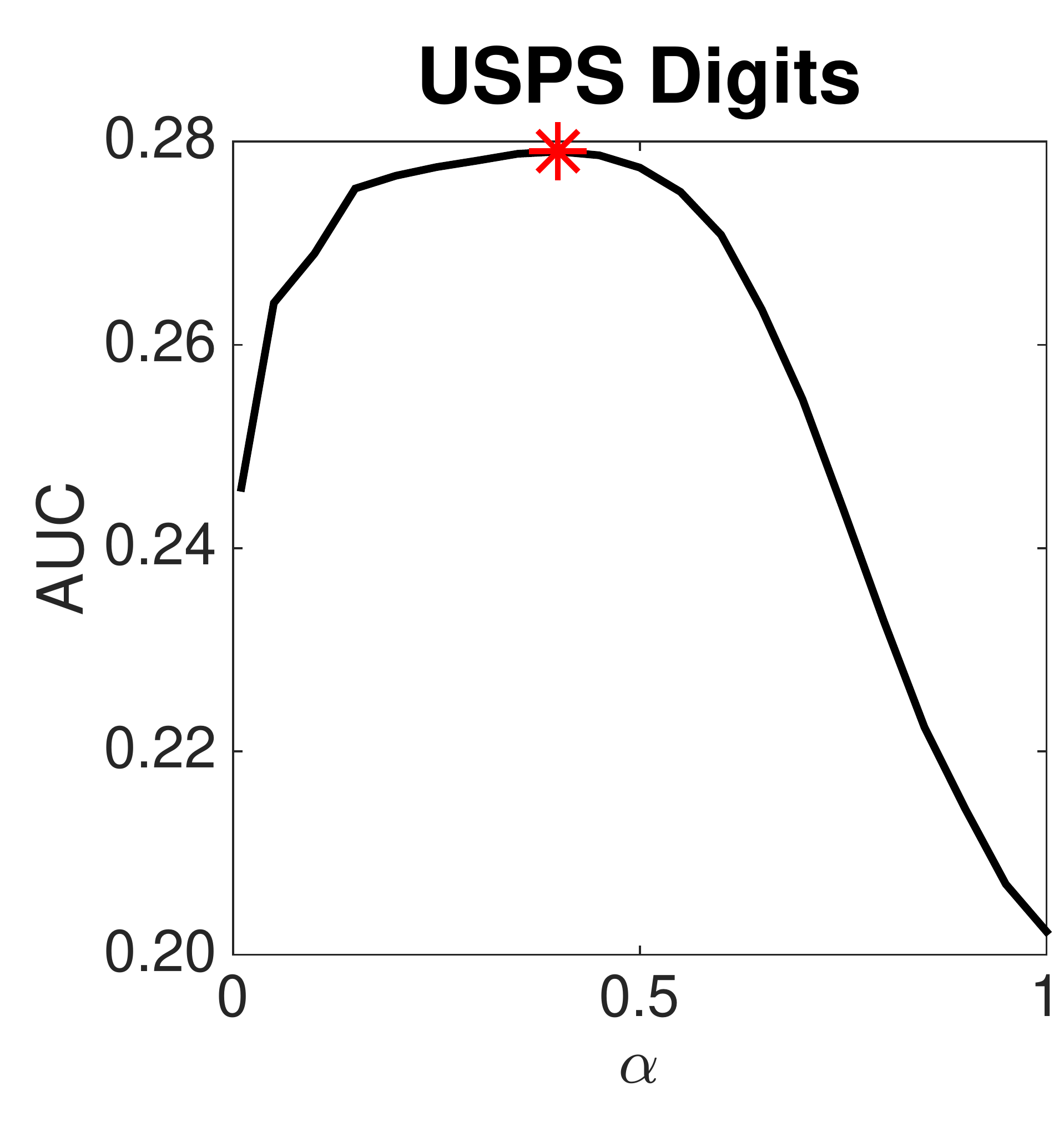}
                \label{fig:auc_usps}}\hfill
        \subfigure{
                 \includegraphics[width=0.23\textwidth]{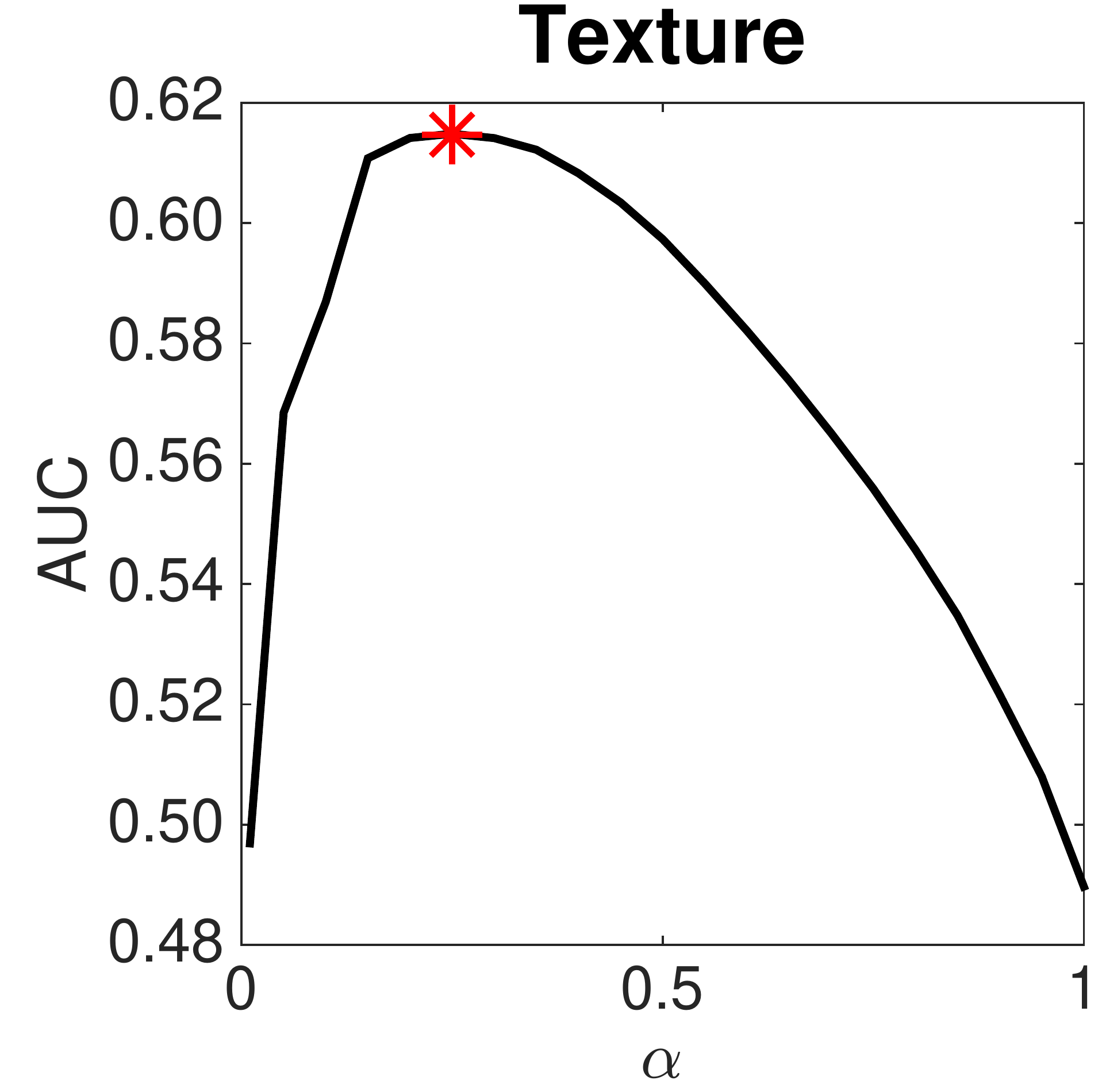}
                \label{fig:auc_text}}\hfill\\\vspace{-1\baselineskip}
         \setcounter{subfigure}{0}
         \subfigure[]{
                \includegraphics[width=0.23\textwidth]{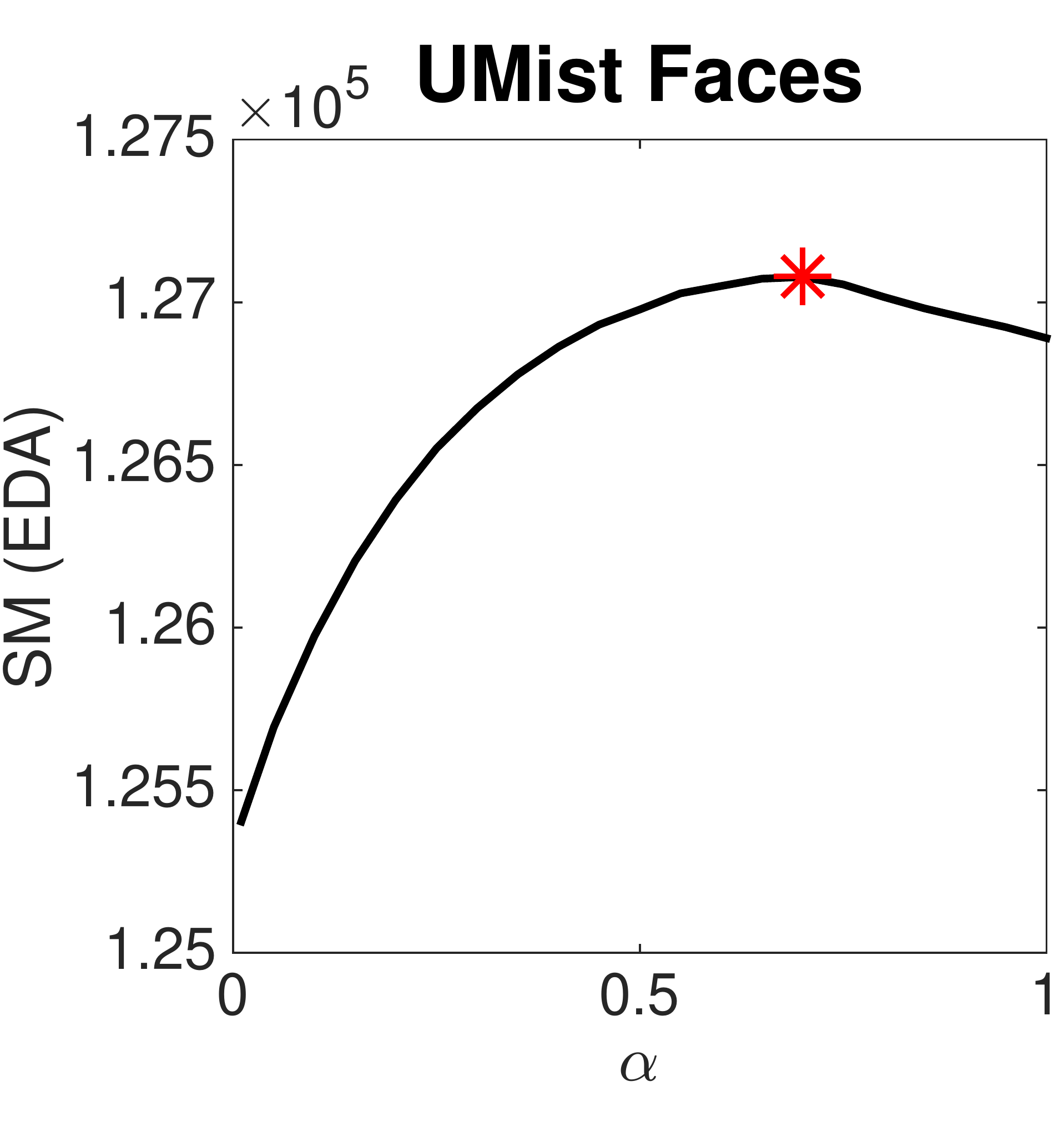}
                \label{fig:eda_ufaces}}\hfill
        \subfigure[]{
                 \includegraphics[width=0.23\textwidth]{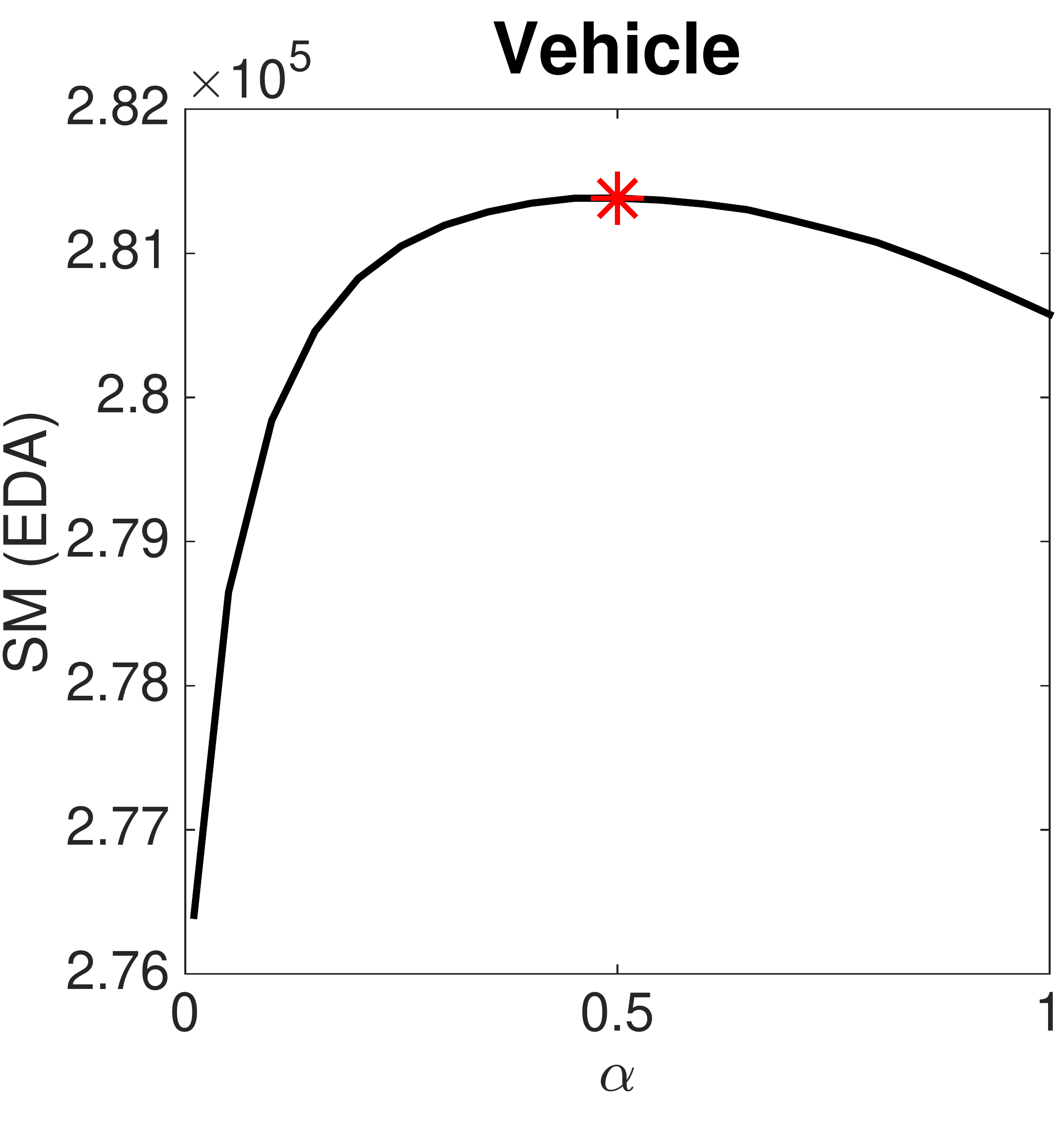}
                \label{fig:eda_vehicle}}\hfill
        \subfigure[]{
                 \includegraphics[width=0.23\textwidth]{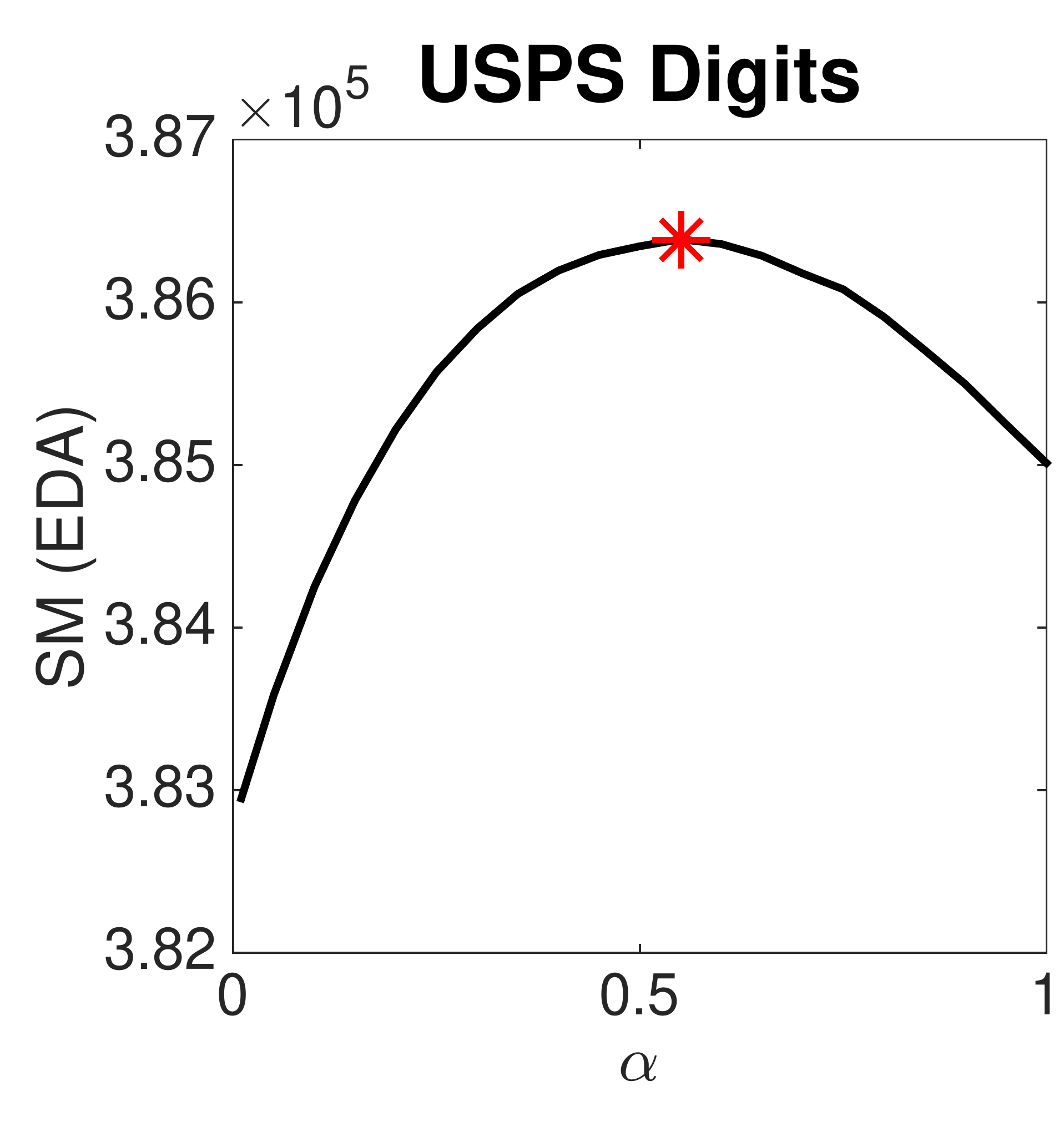}
                \label{fig:eda_usps}}\hfill
        \subfigure[]{
                 \includegraphics[width=0.23\textwidth]{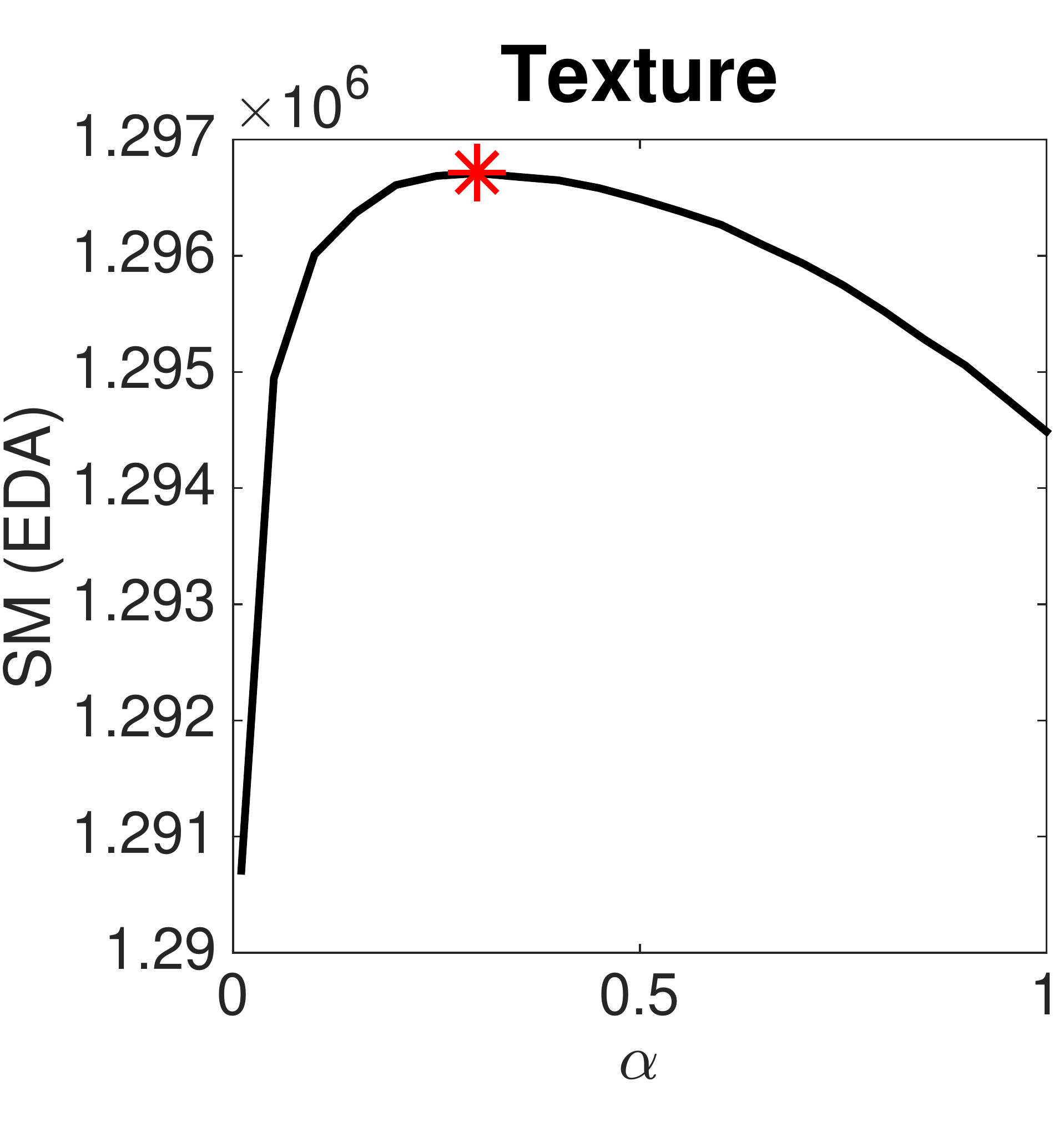}
                \label{fig:eda_text}}\hfill\\
        \caption{Area under ROC curve (first row), and log-likelihood obtained using EDA on (a) \textbf{UMist Faces}, (b) \textbf{Vehicle}, (c) \textbf{USPS Digits}, and (d) \textbf{Texture} datasets.}\label{fig:auc_eda}
\end{figure*}

The results are shown in Table~\ref{tab:results}. 
As can bee seen from the table, {\sf $\alpha$-SNE} method outperforms the other methods on all datasets by means of AUC. Additionally, the value obtained by \textsf{EDA} is exactly the same or very close to the maximum value, obtained by linear search. Note that the estimation of optimal $\alpha$ using \textsf{EDA} becomes more accurate as the size of the dataset increases. This is a result of having more accurate data distribution as more data samples become available. Figure~\ref{fig:auc_eda} illustrates the examples of AUC and log-likelihood curves on different datasets, obtained by varying the $\alpha$ parameter. Clearly, the maximum likelihood estimate of $\alpha$ using \textsf{EDA} and the $\alpha$ value that maximizes the AUC, i.e., that yields the optimal trade-off between precision and recall, coincide in most cases, or at least, are located very close together. This confirms our claim that the optimal value of $\alpha$ obtained by \textsf{EDA} is the one that yields a visualization that represents the data best.

\begin{figure*}[t!]
        \centering
        \subfigure[]{
                \includegraphics[width=0.22\textwidth]{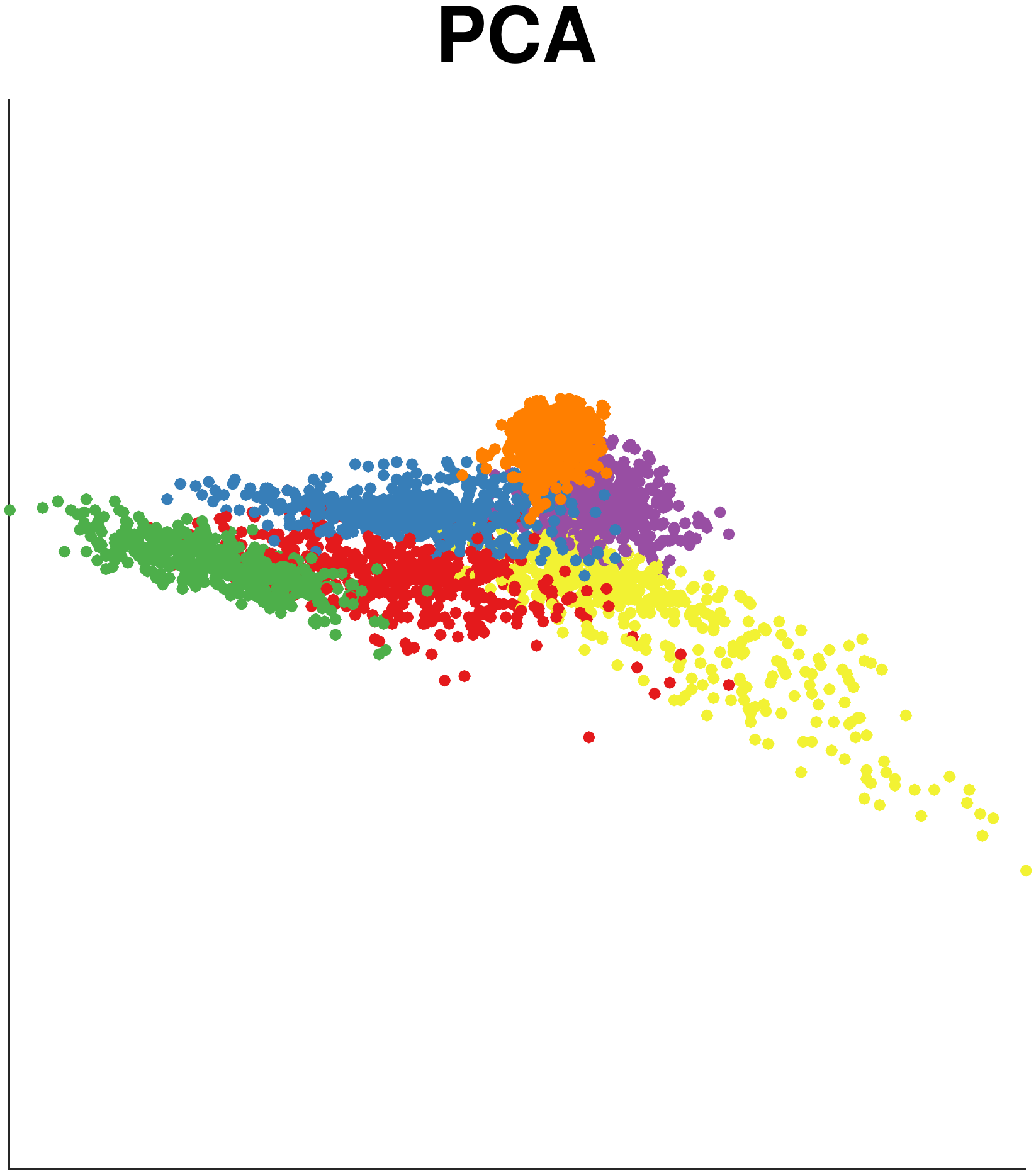}
                \label{fig:pca_text}}
        \subfigure[]{
                 \includegraphics[width=0.22\textwidth]{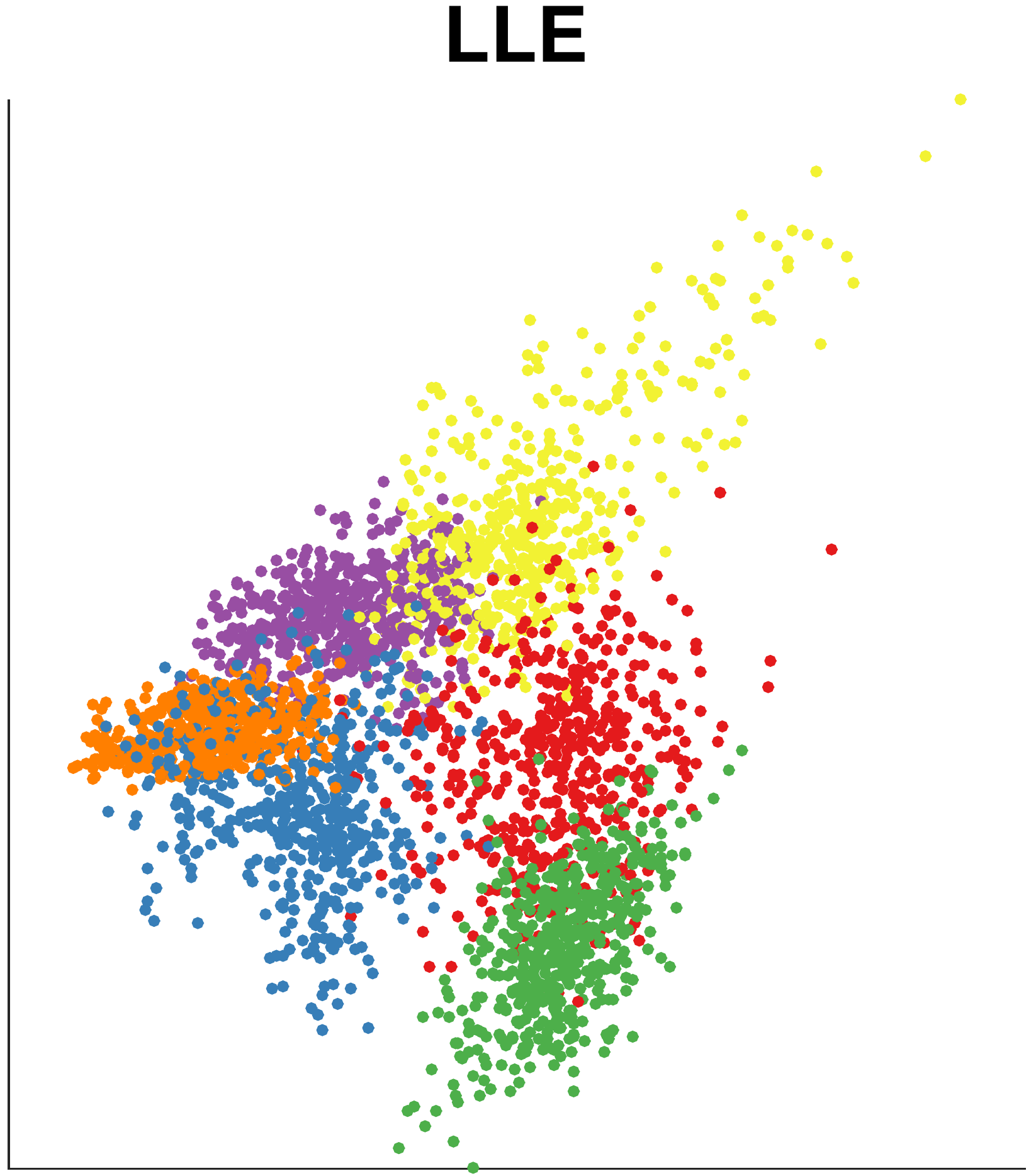}
                \label{fig:lle_text}}
        \subfigure[]{
                 \includegraphics[width=0.22\textwidth]{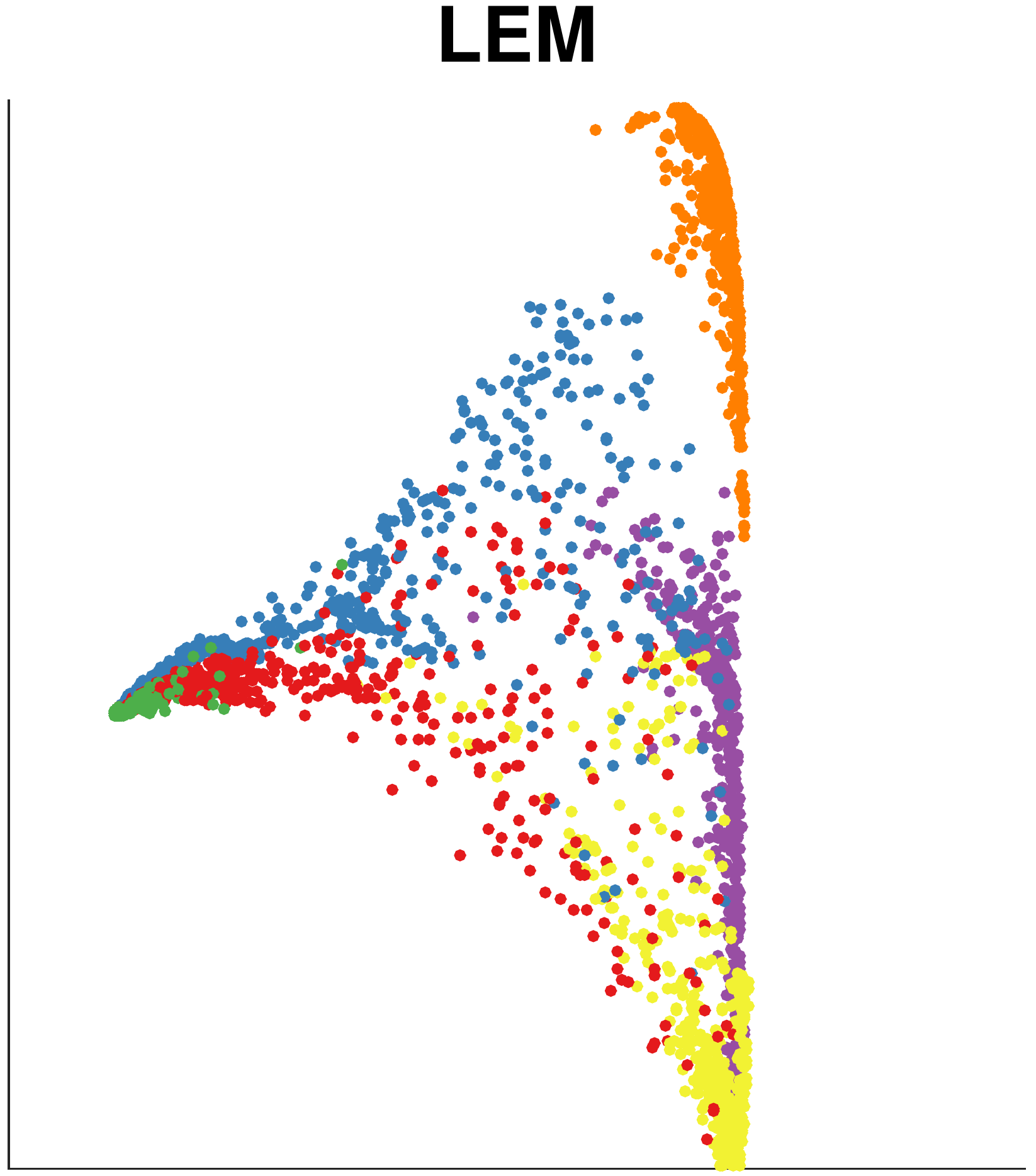}
                \label{fig:lem_text}}
        \subfigure[]{
                 \includegraphics[width=0.22\textwidth]{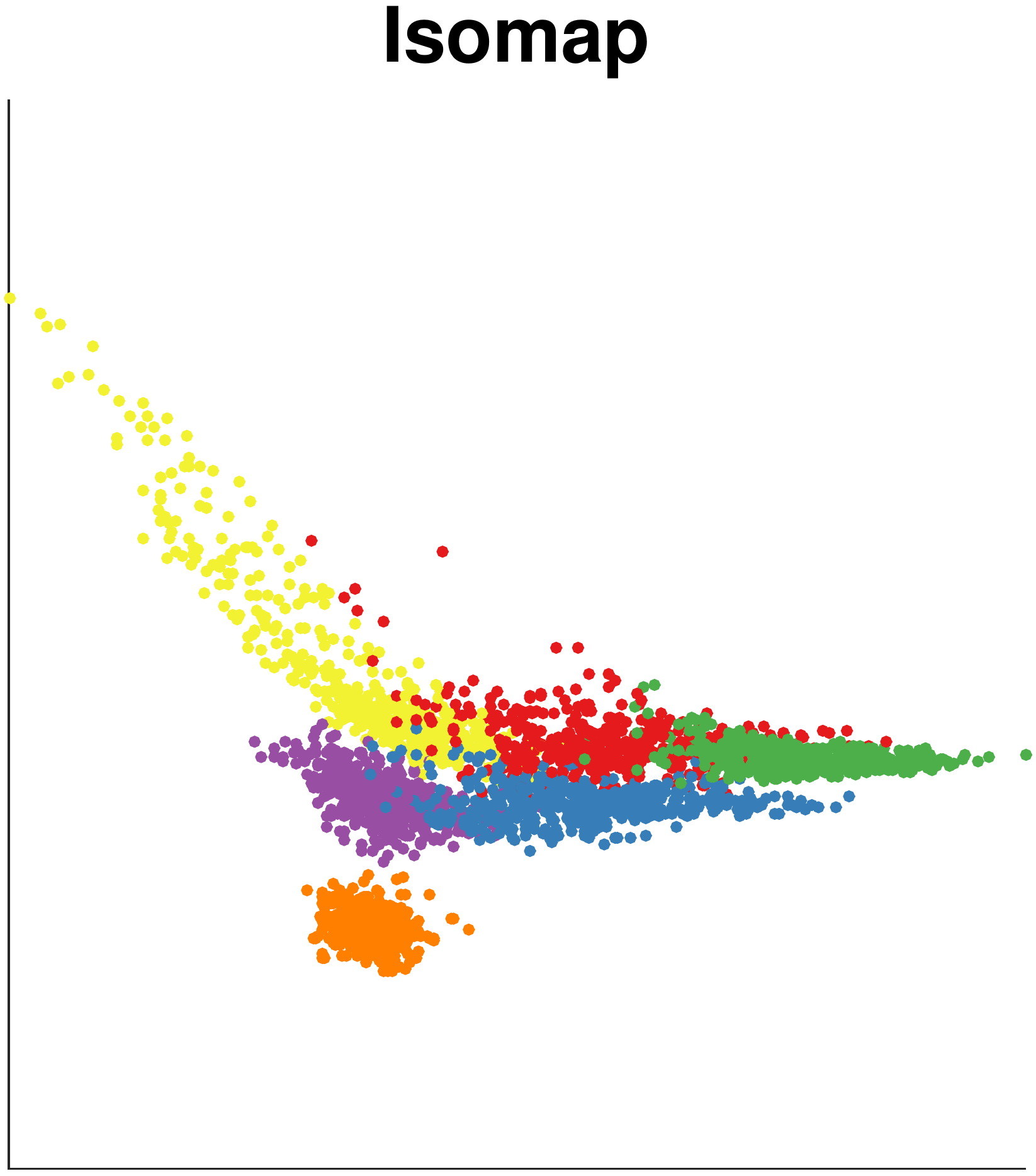}
                \label{fig:isomap_text}}\\\vspace{-2\baselineskip}
         \subfigure[]{
                \includegraphics[width=0.22\textwidth]{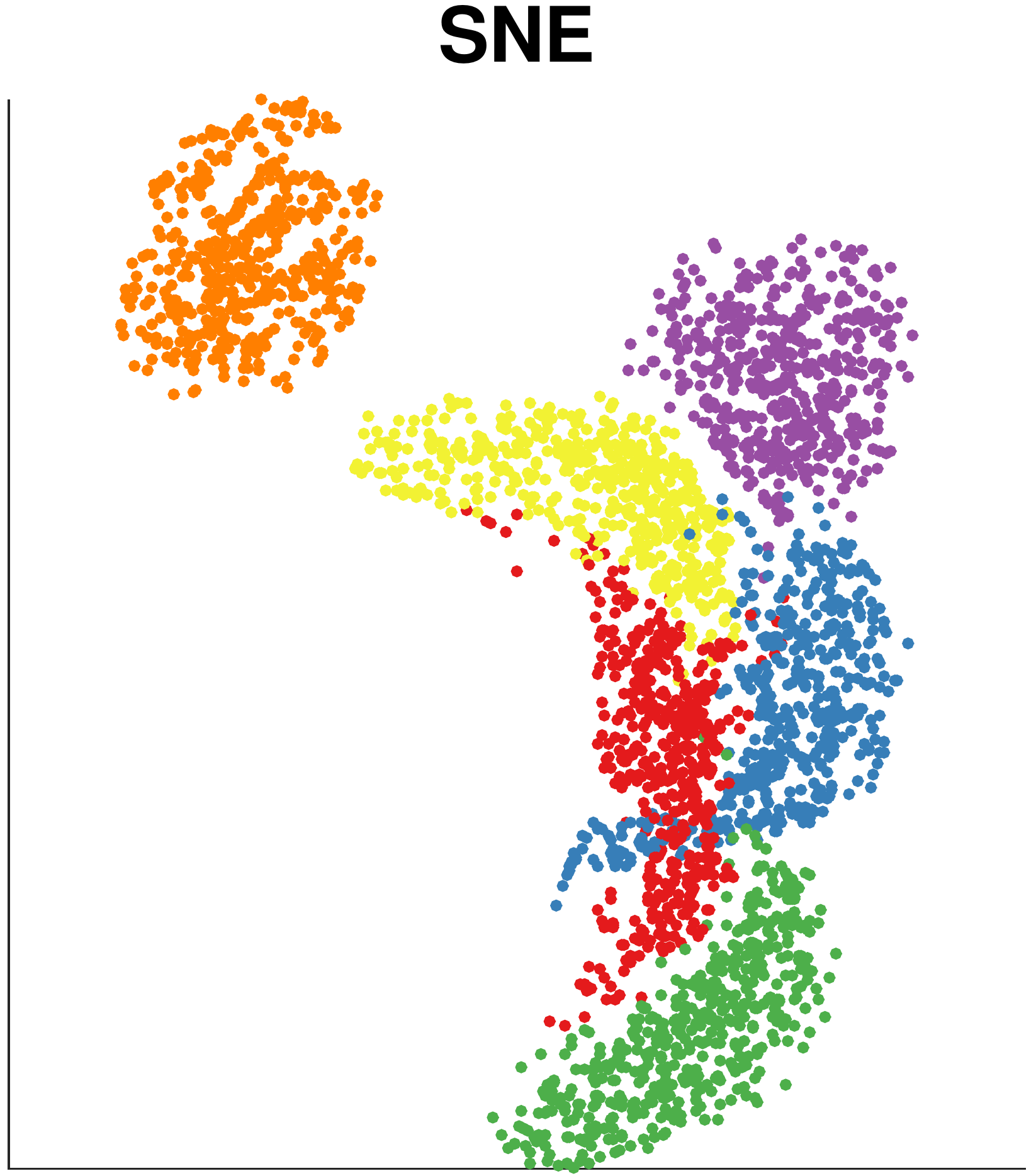}
                \label{fig:sne_text}}
        \subfigure[]{
                 \includegraphics[width=0.22\textwidth]{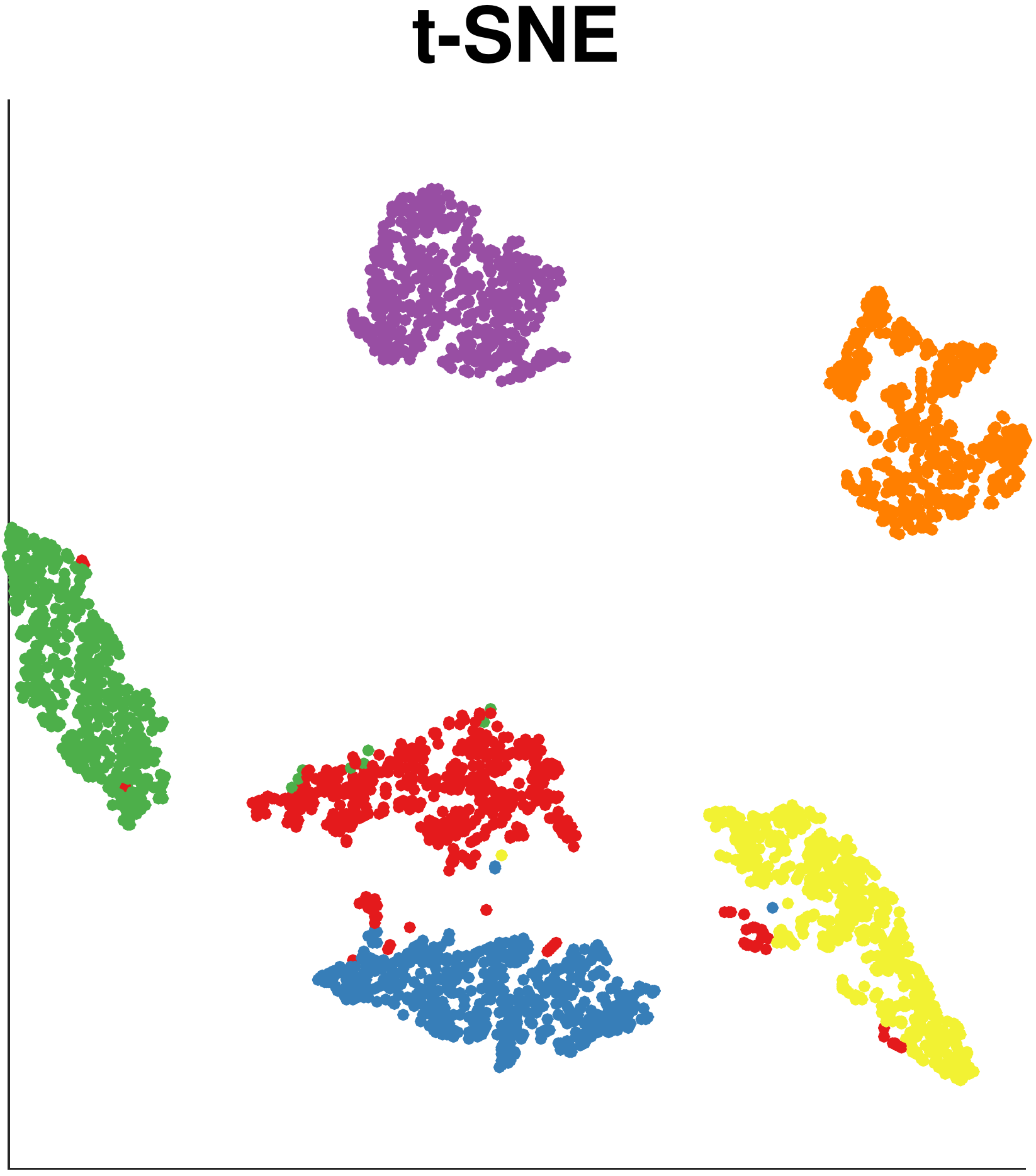}
                \label{fig:tsne_text}}
        \subfigure[]{
                 \includegraphics[width=0.22\textwidth]{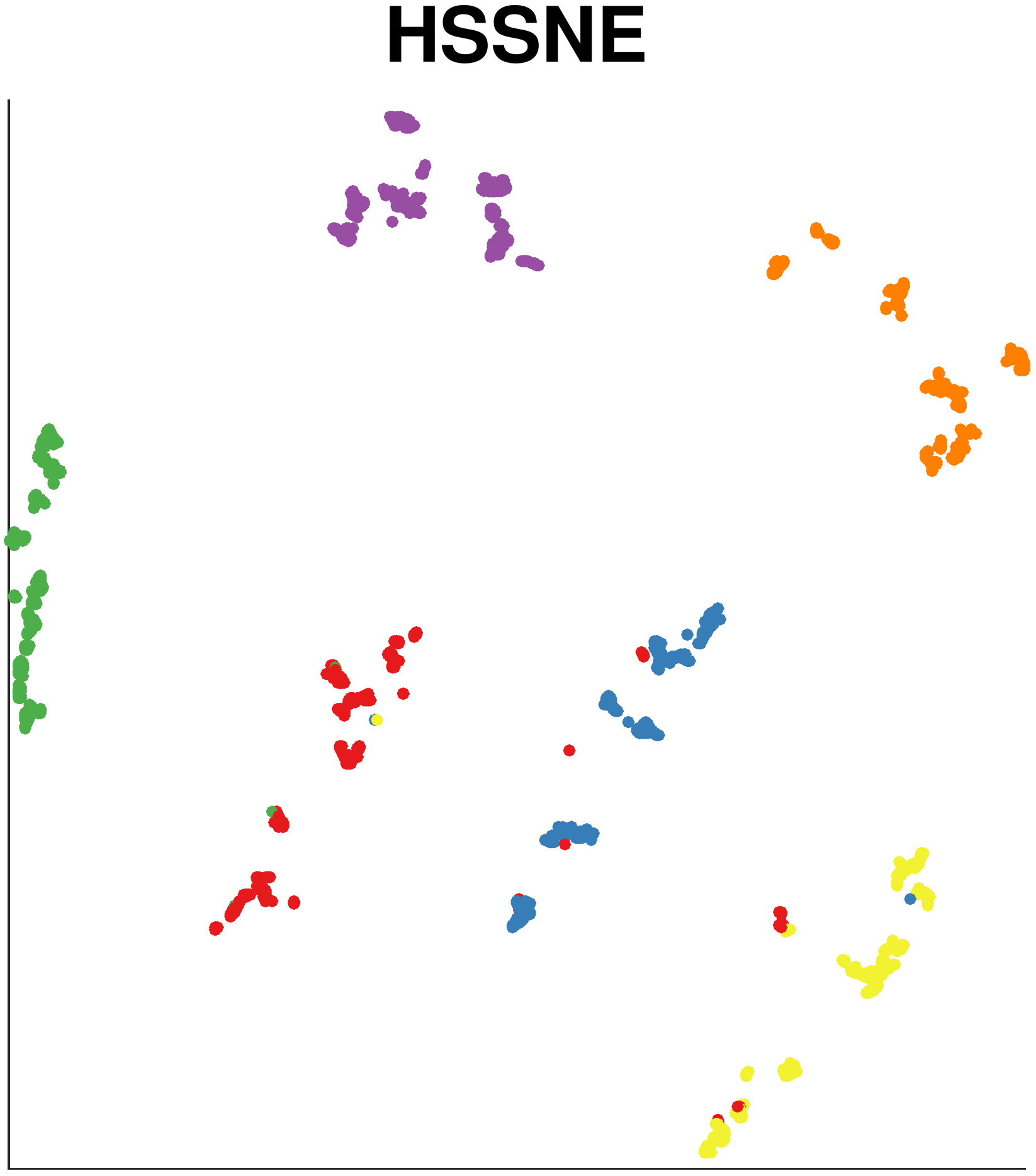}
                \label{fig:hssne_text}}
        \subfigure[]{
                 \includegraphics[width=0.22\textwidth]{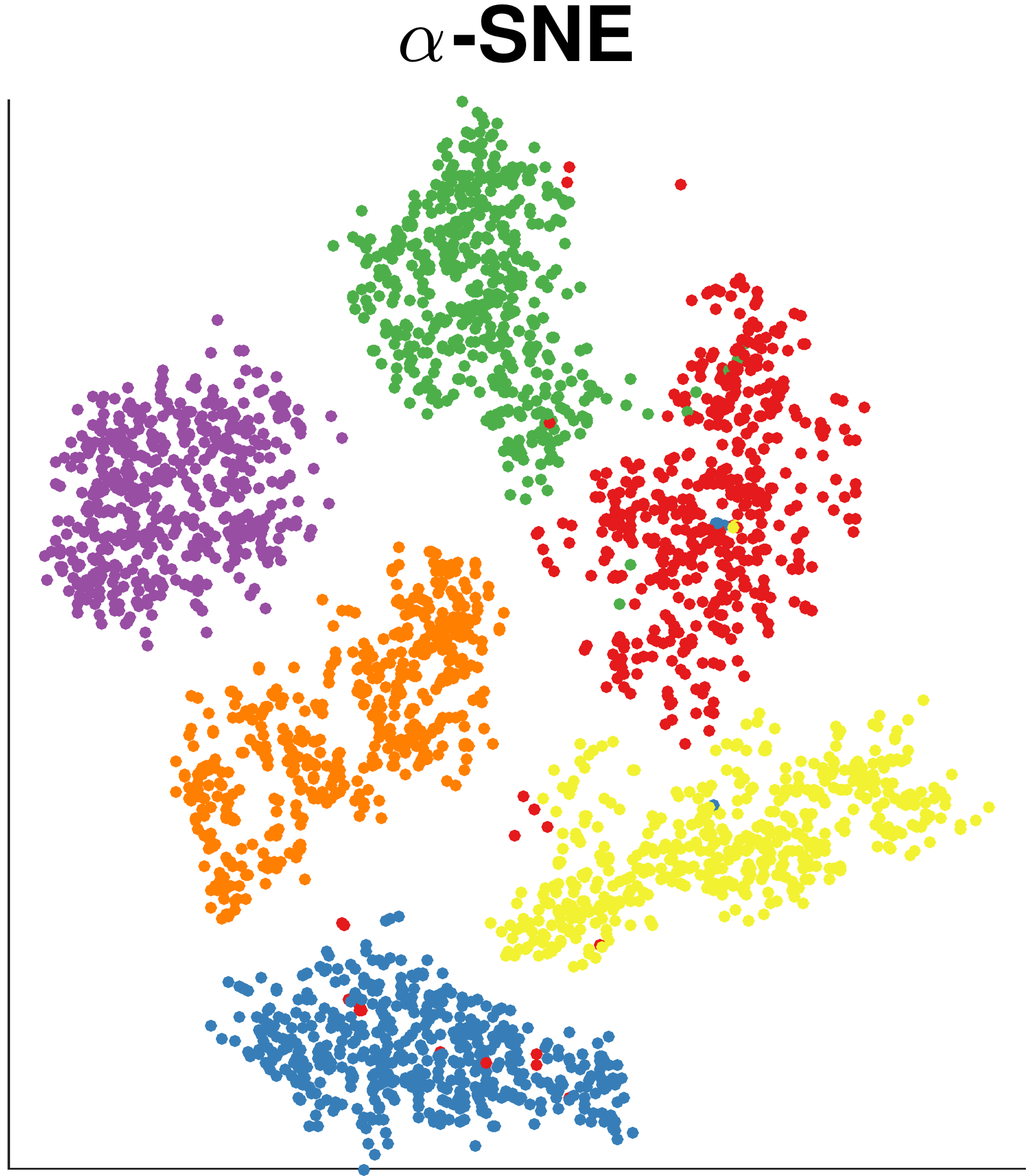}
                \label{fig:asne_text}}\\
        \caption{Visualization od the \textbf{Texture} dataset using different dimensionality reduction methods: (a) \textsf{PCA}, (b) \textsf{LLE}, (c) \textsf{LEM}, (d) \textsf{Isomap}, (e) \textsf{SNE}, (f) \textsf{t-SNE}, (g) \textsf{HSSNE}, and (h) \textsf{$\alpha$-SNE}.}\label{fig:visualizations}
\end{figure*}

Finally, Figure~\ref{fig:visualizations} shows the visualization of the \textbf{Texture} dataset using different dimensionality reduction methods. As can be seen, the clusters are well separated only using \textsf{t-SNE}, \textsf{HSSNE}, and \textsf{$\alpha$-SNE} methods. However, the result of \textsf{t-SNE} is somehow misleading as the clusters are excessively over-separated. This can be seen by increasing the tail-heaviness parameter from $\omega = 1$ in \textsf{t-SNE} to $\omega = 2$ in \textsf{HSSNE}, where the clusters are even more squeezed. This is in contrast with reality where the clusters are distinguished but located fairly close to each other. The result obtained by \textsf{$\alpha$-SNE} illustrates this property by maximizing the AUC of the visualization. As the final remark, the visualization obtained from \textsf{NeRV} is visually similar to the one obtained from \textsf{$\alpha$-SNE}, however, yields a lower AUC value.

\section{Conclusions and Future Work}
\label{sec:conclusion}

We proposed the {\sf $\alpha$-SNE} method to obtain a faithful representation of the data on a two-dimensional screen. We showed that the minimization of the cost function corresponds to maximizing the geometric mean of the precision and recall of the visualization. We also proposed a statistical framework to estimate the optimal $\alpha$ parameter for the visualization, purely based on the data. By an extensive set of experiments, we showed that our proposed method outperforms the previous dimensionality reduction methods by means of the quality of the visualization. Additionally, the experiments verify our claim that the optimal value of $\alpha$ under the \textsf{EDA} distribution yields the optimal trade-off between precision and recall.

As future work, we would like to extend our method to incorporate heavy-tailed distributions in the low-dimensional space. The generalization of optimization of the \textsf{EDA} to incorporate the joint distribution over $\alpha$ and tail-heaviness parameter is left to future work.

\appendix

\section*{Appendix A. Proof of the Connection Between $\alpha$-SNE Cost Function and Precision and Recall}
\label{app:prec_recall}

We consider the binary neighborhood model in both the input space and the low-dimensional embedding, similar to~\cite{nerv}. By binary neighborhood, we assume that each point has a fixed number of equally relevant neighbors in both the input and the embedding. Under this model, the probabilistic neighborhood models in the input space and the embedding are defined as
\begin{equation}
\label{eq:p_bin}
p_{ij} = \left\{ 
  \begin{array}{l l l}
  a_i \equiv \frac{1-\delta}{n_i}, & &\text{if } j \in P_i\\
  b_i \equiv \frac{\delta}{N-n_i+1}, & &\text{otherwise}
   \end{array} \right. \,, \qquad q_{ij} = \left\{ 
  \begin{array}{l l l}
  c_i \equiv \frac{1-\delta}{r_i}, & &\text{if } j \in Q_i\\
  d_i \equiv \frac{\delta}{N-r_i+1}, & &\text{otherwise}
   \end{array} \right. \,.
\end{equation}
where $N$ is the number of the points and $0\leq \delta \ll 0.5$ is a small constant, assigning a small portion of the probability mass to irrelevant points.

We now show that minimizing the $\alpha$-divergence between the probabilities in the input space and the embedding is equivalent to maximizing the geometric mean of precision and recall. We consider the case $0 < \alpha < 1$. Similar results hold for $\alpha = 0, 1$ (see~\cite{nerv}). Using the fact that $\sum_i p_i = \sum_i q_i = 1$ and ignoring the constant factors, the $\alpha$-divergent cost function to be minimized becomes
\begin{equation}
\label{eq:cost_min}
\begin{split}
D_\alpha (\p_i \Vert \q_i) \propto 1 & - \sum_{j \in P_i, j \in Q_i} a_i^{\alpha} c_i^{1-\alpha} - \sum_{j \in P_i, j \notin Q_i} a_i^{\alpha}d_i^{1-\alpha}\\& - \sum_{j \notin P_i, j \in Q_i} b_i^{\alpha} c_i^{1-\alpha} - \sum_{j \notin P_i, j \notin Q_i} b_i^{\alpha} d_i^{1-\alpha}\\
= 1 - \left(a_i^{\alpha} c_i^{1-\alpha}\right) & N_{\text{TP}} - \left(a_i^{\alpha}d_i^{1-\alpha}\right) N_{\text{miss}} - \left(b_i^{\alpha} c_i^{1-\alpha}\right) N_{\text{FP}} - \left(b_i^{\alpha} d_i^{1-\alpha}\right) N_{\text{TN}}
\end{split}
\end{equation}
Substituting the values in the cost function gives
\begin{equation}
\label{eq:cost_min_reduced}
\begin{split}
D_\alpha (\p_i \Vert \q_i) \propto 1 & - \frac{(1-\delta)}{r_i^{\alpha}k_i^{1-\alpha}} N_{\text{TP}} - \frac{(1-\delta)^\alpha \delta^{1-\alpha}}{r_i^{\alpha}(N- k_i - 1)^{1-\alpha}} N_{\text{miss}}\\
& - \frac{\delta^\alpha(1-\delta)^{1-\alpha}}{(N-r_i-1)^{\alpha}k_i^{1-\alpha}} N_{\text{FP}} - \frac{\delta}{(N- r_i - 1)^{\alpha}(N-k_i-1)^{1-\alpha}} N_{\text{TN}}\, .
\end{split}
\end{equation}
For small values of $\delta$, the first two terms are dominating and the remaining terms become negligible. Therefore, in the limit $\delta \rightarrow 0$, minimizing the cost in~(\ref{eq:cost_min_reduced}) becomes equivalent to maximizing the term 
\begin{equation*}
\text{PR}(\alpha) = \frac{N_{\text{TP}}}{r_i^\alpha k_i^{1-\alpha}}
\end{equation*}
which is the geometric mean of precision and recall, parameterized by $\alpha$. The result is also consistent in the limit points $\alpha = 0$ and $\alpha = 1$ where $\text{PR}(0) = \frac{N_{\text{TP}}}{ k_i}$ and $\text{PR}(1) = \frac{N_{\text{TP}}}{ r_i}$ reduce to recall and precision, respectively. As a direct result of geometric-arithmetic mean inequality, for any values of $\alpha = \lambda$,  $\alpha$-divergence maximizes an upper bound for the convex sum of precision and recall, that is,
\begin{equation}
\lambda \frac{N_{\text{TP}}}{ r_i} + (1-\lambda) \frac{N_{\text{TP}}}{ k_i} \leq \frac{N_{\text{TP}}}{r_i^\alpha k_i^{1-\alpha}}\, .
\end{equation}

\bibliographystyle{splncs}
\bibliography{refs}

\end{document}